\documentstyle[prl,twocolumn,aps]{revtex}
\input{epsf}
\begin{document}
\draft
\twocolumn[\hsize\textwidth\columnwidth\hsize\csname @twocolumnfalse\endcsname
\title{Quantum computer inverting time arrow for macroscopic systems}

\author{B. Georgeot and D. L. Shepelyansky}

\address {Laboratoire de Physique Quantique, UMR 5626 du CNRS, 
Universit\'e Paul Sabatier, F-31062 Toulouse Cedex 4, France}

%\date{\today}
\date{May 30, 2001}

\maketitle

\begin{abstract}
\end{abstract}
%\pacs{PACS numbers: 05.45.+b, 75.10.Jm, 75.10.Nr}
\vskip1pc]

%\begin{multicols}{2}
\narrowtext

% Introduction
{ \bf A legend tells \cite{mayer} that once Loschmidt asked Boltzmann on what
happens to his statistical theory if one inverts the velocities
of all particles, so that, due to the reversibility
of Newton's equations, they return from the equilibrium
to a nonequilibrium initial state.  Boltzmann only replied ``then go
and invert them''.  This problem of the relationship
between the microscopic and macroscopic descriptions of the physical world and
time-reversibility has been hotly debated from the XIX${th}$ century up to 
nowadays \cite{boltzmann,landauer,bennett,zurek,lebowitz,prigogine,ruelle,hoover}.  
At present, no modern computer is able to perform
Boltzmann's demand for a macroscopic number of particles.  In addition,
dynamical chaos \cite{arnold,sinai,chirikov,lichtenberg}
implies exponential growth of any imprecision in
the inversion that leads to practical irreversibility.  Here we
show that a quantum computer \cite{divi,ekert,steane,kane}
composed of a few tens of qubits, 
and operating even  with moderate precision, can perform Boltzmann's
demand for a macroscopic number of classical particles. 
Thus, even in the regime of dynamical chaos, a realistic 
quantum computer allows to rebuild a specific initial 
distribution  from a macroscopic state given by thermodynamic  laws.
%Thus, even in the regime
%of Fokker-Planck diffusion generated by  dynamical chaos, a realistic 
%quantum computer allows to invert the diffusion and rebuild a specific 
%initial state from macroscopic statistical equilibrium. 
}

To study the relations between microscopic deterministic classical dynamics,
macroscopic thermodynamic laws and quantum computation, we choose 
a simple area-preserving map: 
\begin{equation}
\label{catmap}
\bar{y}=y+x \; \mbox{(mod} \;\mbox{L)}\;\;, \;\; 
\bar{x}=x+\bar{y} \;\mbox{(mod} 
\;\mbox{1)}\;.
\end{equation}
Here the first equation can be seen as a kick which changes the momentum 
$y$ of a particle, while the second one corresponds to a free phase rotation
in the interval $-0.5\leq x < 0.5$; bars mark the new values of the variables.
The map dynamics takes place on a torus of integer length $L$ in the 
$y$ direction.  For $L=1$ this map reduces to the well-known Arnold cat map
\cite{arnold}, which describes a fully chaotic dynamics with positive
Kolmogorov-Sinai entropy $h\approx 0.96$.  As a result, the dynamics is
characterized by exponential divergence of nearby trajectories, so that
any small error $\epsilon$ (for example round-off error) 
grows exponentially with time, and reversibility of a trajectory is lost
after $t_E \approx |\ln \epsilon | /h$ map iterations.  
For $\epsilon \sim 10^{-8}$ comparable to ordinary precision
of the Pentium III, this time scale is rather short ($t_E \approx 20$).  For 
$ L \gg 1$ chaos leads to the diffusive spreading of particles in momentum,
which is well described by the Fokker-Planck  equation:
\begin{equation}
\label{fokker}
 \partial w(y,t)/ \partial t = D/2\; \; \partial^2 w(y,t)/ \partial^2 y,
\end{equation}
where the diffusion coefficient $D \approx <x^2>= 1/12$.
Thus after a time $t \gg 1/h$ an initial distribution of particles in 
(\ref{catmap}) evolves towards a Gaussian statistical distribution
$w(y,t)=w_g(y,t)=\exp(-(y-y_0)^2/(2Dt))/\sqrt{2D\pi t}$
with $<y^2> = Dt +y_0^2$, where $y_0= <y>$ at $t=0$.  
On a finite torus this diffusive process 
relaxes to a homogeneous distribution in $y$
after a time $t_D \approx L^2/D $. 

For the case $L=1$ it was shown that a quantum computer
can simulate a discretized version of this map with exponential efficiency
\cite{ascat}.  Here we show that for $L \gg 1$
a similar quantum algorithm enables to 
simulate the evolution of a macroscopic number of classical particles 
which is governed by the thermodynamic diffusion law.  To perform this
evolution on a lattice of size $LN^2$
(with $N=2^{n_q}$ and $L=2^{n_{q'}-n_q}$)
this algorithm uses three quantum registers. 
The first one with $n_q$ qubits
holds the values of the coordinate $x$ ($x_i= -0.5+i/N, i=0,...,N-1$),
the second one with $n_{q'}$ qubits holds the $y$ coordinates 
($y_j= -L/2+j/N, j=0,...,LN-1$) and the last one with $n_{q'}-1$ 
qubits is used as a workspace.  The first two registers describe the 
discretized classical phase space with $L$ cells and $N^2$ points per
cell.  In this way, the initial positions of $N_d \sim N^2$  particles 
can be represented
by one quantum state $\sum_{i,j} a_{ij} |x_i> |y_j>|0>$, where 
$a_{i,j}= 0$ or $1/\sqrt{N_d}$.  The quantum algorithm is based on
modular additions performed in a way similar to the one described in
\cite{barrenco}, through Toffoli and controlled-not gates (CNOT).  
It requires $10n_q + 6n_{q'}-17$ gate operations per map iteration, 
in contrast to $O(2^{2n_q})$ operations
for the classical algorithm. The time inversion 
is also realized by $8n_q +4n_{q'} -13$ gate
operations which effectively change $y$ into $ -y$ half-way between kicks 
\cite{ascat}\cite{pgm1}.
In this way, the quantum computer acts in a way similar to Maxwell's demon 
\cite{maxwell,leff} who 
reverses the velocity of each individual particle.

A perfect quantum computer simulates exactly the map (\ref{catmap}), but 
a realistic physical system always has some imperfections which can destroy 
time-reversibility.  For a classical computer, e. g. Pentium III 
iterating map (\ref{catmap}), round-off errors of amplitude $\epsilon$
destroy the time-reversibility of the map dynamics after $t_E$ iterations.
This fact is illustrated on Fig.1 where it is assumed that the demon
inverts the velocities of all trajectories after $t_r=35$ iterations with
a precision $\epsilon$.  After that, the macroscopic distribution 
starts to return back but after $t_E\approx |\ln \epsilon | /h $ 
iterations the errors
become too large and the diffusion process restarts again.
In contrast to that, on a quantum computer round-off errors can be
decreased enormously since the size of the registers grows only
linearly with $n_q$ and the computation is exponentially efficient.
However, a quantum computer has its own natural errors which can be
viewed as imprecisions of amplitude $\epsilon$ in the gate operations. 
The comparison of the two types of errors natural for classical
and quantum computers is displayed on Fig.1.  It shows that the
quantum computation with precision $\epsilon=0.01$ in each gate at each
map iteration is able to reverse effectively the diffusion process up
to the initial state.  That is in striking contrast with the irreversibility
of the classical computation with a round-off error of amplitude
$\epsilon=10^{-8}$ made only once at $t=t_r$ when the demon acts.
In this way, the quantum computer succeeds to reverse the thermodynamic
diffusive process with enormous number of particles.  Indeed, at the moment
of inversion $t_r$, the distribution of particles is a Gaussian of width
$\sigma= \sqrt{2Dt_r}$ in agreement with the solution of (\ref{fokker}),
as is shown in Fig.2.

\begin{figure}
\epsfxsize=3.4in
\epsfysize=2.6in
\epsffile{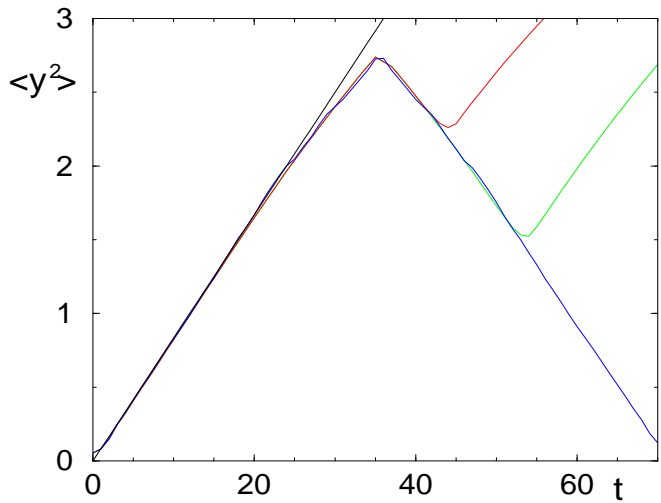}
\vglue 0.2cm
%\medskip
\caption{Diffusive growth of the second moment $<y^2>$ 
of the distribution $w(y,t)$ generated by the map (\ref{catmap})
with $L=8$,
simulated on a classical (Pentium III)
and quantum (``Quantium I'') computers. At $t=t_r=35$ one
inverts all velocities.  For Pentium III inversion is done with
precision $\epsilon=10^{-4}$ (red line) and  $\epsilon=10^{-8}$
(green line); $10^6$ orbits are simulated, initially distributed
inside the demon image (see Fig. 3).  For Quantium I, the computation
is done with $26$ qubits ($n_q=7, n_{q'}=10$)(blue line);  
each quantum gate
operates with imperfections of amplitude $\epsilon=0.01$ (unitary rotation 
on a random angle of this amplitude).  The black straight line shows the
theoretical macroscopic diffusion with $D=1/12$.}
\label{fig1}
\end{figure}

Fig.3 shows explicitly the distribution in phase space at different moments of 
time.  The initial distribution mimics a demon, which at $t=t_r$ is transformed
to a statistical homogeneous distribution in the $x$ direction, with a smooth
variation in $y$ described by (\ref{fokker}).  The quantum computer operating
with $1\%$ accuracy is able to recover the initial image with good precision,
whereas the classical computer with round-off errors $10^{-8}$ completely
fails to reproduce it.  The striking difference between the two final 
distributions at $t=t_{2r}$ generated by the two computers 
can be easily detected from a polynomial number of measurements.

\begin{figure}
\epsfxsize=3.4in
\epsfysize=2.6in
\epsffile{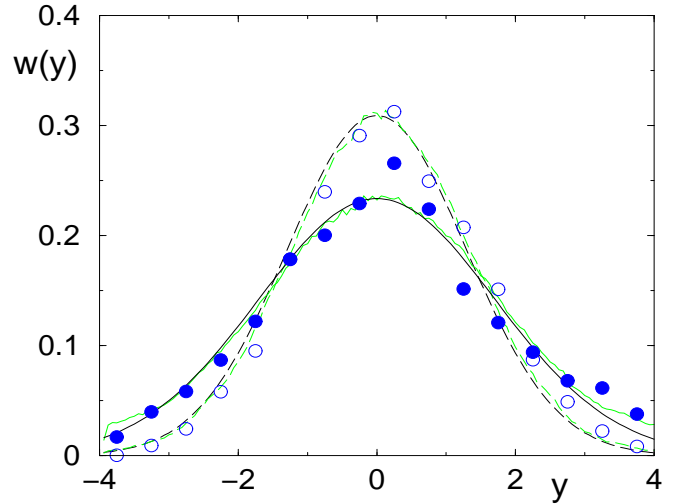}
\vglue 0.2cm
%\medskip
\caption{Distribution of particles in $y$ for map (\ref{catmap})
simulated by Pentium III (green lines) and Quantium I (blue symbols)
for the case of Fig.1,
at $t=20$ (dashed lines and open circles) and $t=t_r=35$ (full lines
and filled circles).  Black lines show the theoretical 
solution of the Fokker-Planck equation (\ref{fokker}).}
\label{fig2}
\end{figure}

The previous  results are supported by the data for 
the fidelity $f(t)$ defined as 
the projection of the quantum state in presence of gate imperfections
on the exact state without imperfections.  For $f=1$ both states coincide,
whereas for $f \ll 1$ both distributions are completely different.
The results on Fig.4 show that $f(t)$ smoothly decreases with number of 
iterations $t$ even if classical dynamics is exponentially unstable.
The probability of transition from the exact state to other states
induced by imperfections can be estimated as of the order of $\epsilon^2$.
Hence, since imperfections in each gate are assumed to be uncorrelated,
$f(t)$ should drop by $n_q\epsilon^2$ at each map iteration 
(for $n_q \sim n_{q'}$).  This determines a time scale 
\begin{equation}
\label{timescale}
t_f \approx C/(n_q \epsilon^2)
\end{equation}
on which the fidelity of quantum computation for the algorithm is reasonable
($f(t_f)=0.5)$),
even in absence of error correction.  This scaling is in agreement with
the data in Fig.4 (see also \cite{ascat})  which give the numerical
factor $C\approx 0.5$.  This is in sharp contrast with
classical errors for which computation of trajectories remains correct 
only up to a time scale $t_E \approx |\ln \epsilon|/h$.  It is interesting to
note that the situation is similar to the time evolution of a physical
system in the regime 
of quantum chaos, which is stable against small quantum errors even though 
the underlying classical dynamics is chaotic \cite{dima}. 

\begin{figure}
\epsfxsize=2.8in
\epsfysize=7.in
\epsffile{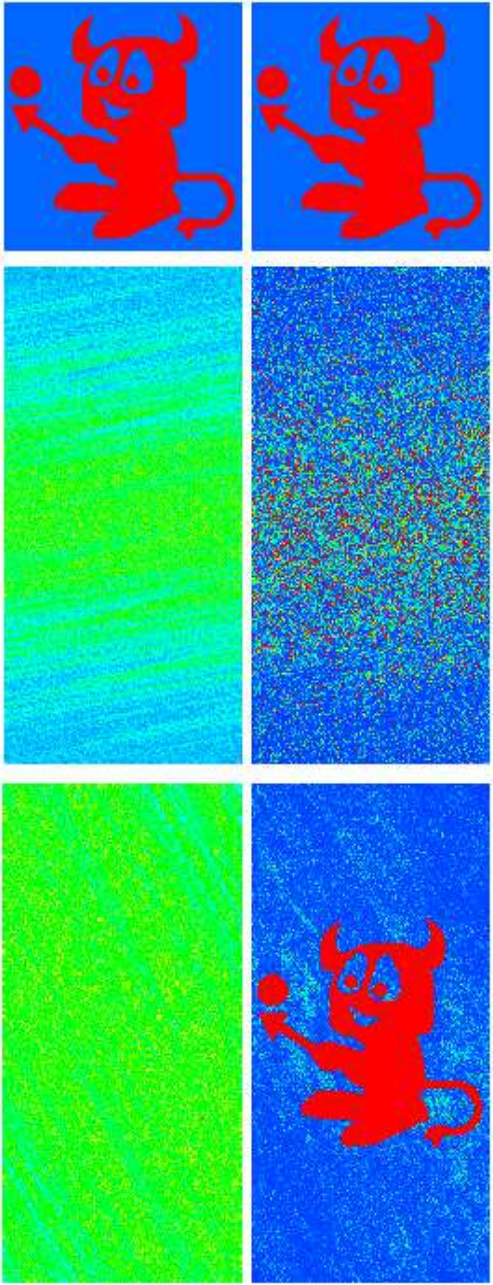}
\vglue 0.2cm
%\medskip
\caption{Evolution of a demon image through map (\ref{catmap}).
Left column shows the simulation on  Pentium III, right column
on Quantium I.  Top: initial distribution in the central cell 
($-0.5\leq x,y <0.5$).  Middle: distribution at $t=t_r=35$
in the whole phase space ($-0.5\leq x <0.5, -4\leq y <4$).
Bottom: distribution at $t=2t_r=70$ in the two central cells
($-0.5\leq x <0.5, -1\leq y <1$). The time-inversion is made 
at $t_r=35$, with accuracy $\epsilon=10^{-8}$ for Pentium III (error
is done only at $t_r$), and with accuracy $\epsilon=0.01$ for Quantium I
(error is done at each gate operation). Color marks the density of
particles/probability, from blue (minimal) to red (maximal value). Here
as in Fig.1 $n_q=7, n_{q'}=10$, with in total $26$ qubits used for Quantium I;
for Pentium III, $10^{6}$ orbits are simulated.}
\label{fig3}
\end{figure}

\begin{figure}
\epsfxsize=3.4in
\epsfysize=2.6in
\epsffile{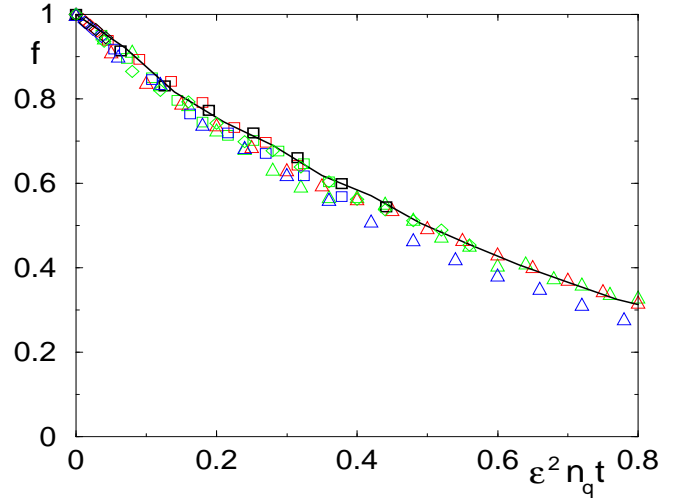}
\vglue 0.2cm
%\medskip
\caption{Fidelity $f$ as a function of $\epsilon^2 n_q t$ for
the quantum simulation of the map (\ref{catmap})
on Quantium I.
Here $L=8$ and $n_q=7$, (black line and symbols), $n_q=6$ (blue symbols),
$n_q=5$ (red symbols),  $n_q=4$ (green symbols)and 
$\epsilon= 10^{-2}$ (diamonds), $\epsilon= 3.10^{-2}$ (squares) and 
$\epsilon= 10^{-1}$ (triangles up).
Black line shows $n_q=7$, $\epsilon= 10^{-1}$.}
\label{fig4}
\end{figure}

The relation (\ref{timescale}) implies that a quantum computer operating
with realistic accuracy can 
invert velocities of all particles at a given moment of time, so that 
a specific initial state is reliably reproduced from a statistical distribution
described by a diffusive process. 
Such a simulation for a macroscopic number of particles $N_d$ can be 
performed with few tens of qubits.  For example, for $N_d=6.022 \times 10^{23}$
(Avogadro's number) the simulation with $L=8$ requires only $125$ qubits.
Moreover, according to (\ref{timescale}), the accurate simulation of
such an enormously large number of particles remains reliable ($f(t)>0.5$)
up to a time $t \approx 150$ with a modest gate accuracy $\epsilon= 0.01$.
Such a computation is far out of reach of any modern supercomputer, and 
clearly shows the power of quantum computers.  
It also opens interesting perspectives
for cryptography since an initial image can be coded
in a thermodynamic distribution with very large entropy
and then reliably recovered.
Thus quantum computers
open new possibilities for the investigation of  
the relations between microscopic
deterministic dynamics and macroscopic thermodynamic laws.

{\bf Acknowledgments:}  We thank the IDRIS in Orsay and CalMiP in Toulouse for 
access to their supercomputers, which were used to simulate Quantium I.
This work was supported in part by the NSA
and ARDA under ARO contract No. DAAD19-01-1-0553 and 
also by the EC  RTN contract HPRN-CT-2000-0156.

Correspondence should be addressed to D.L.Shepelyansky 
(dima@irsamc.ups-tlse.fr).

\vskip -0.5cm
\begin{thebibliography}{99}
\bibitem{mayer} Mayer, J.~E. and Goeppert-Mayer, M. Statistical mechanics.
          (John Wiley \& Sons, N.Y., 1977).
\bibitem{boltzmann} Boltzmann, L. Vorlesungen \"uber Gastheorie, 2 vols.
         (Barth, Leipzig, 1896, 1898) [English translation
         Lectures on gas theory (Cambridge University, London, 1964)].
\bibitem{landauer} Landauer, R. Irreversibility and heat generation
        in the computing process. {\em IBM J. Res. Dev.} {\bf 5}, 183 (1961).
\bibitem{bennett} Bennett, C. H. The thermodynamics of computation - a review.
                  {\em Int. J. Theor. Phys.} {\bf 21}, 905 (1982).
\bibitem{zurek} Zurek, W.~H. Reversibility and stability of information 
processing systems. {\em Phys. Rev. Lett.} {\bf 53}, 391 (1984).
\bibitem{lebowitz} Lebowitz, J. L. Microscopic origins of irreversible
macroscopic behavior. {\em Physica A} {\bf 263}, 516 (1999).
\bibitem{prigogine} Prigogine, I. Laws of nature, probability
and time symmetry breaking. {\em Physica A} {\bf 263}, 528 (1999).
\bibitem{ruelle} Ruelle, D. Gaps and new ideas in our understanding
of nonequilibrium. {\em Physica A} {\bf 263}, 540 (1999).
\bibitem{hoover}  Hoover, W.~G. Time reversibility, computer simulation,
        and chaos. (World Scientific, Singapore, 1999).
\bibitem{arnold}Arnold, V. and Avez, A. Ergodic problems in 
        classical mechanics. (Benjamin, N. Y., 1968).
\bibitem{sinai} Kornfeld, I. P., Fomin, S. V. and Sinai, Ya.~G. 
        Ergodic theory. (Springer, N. Y., 1982).
\bibitem{chirikov} Chirikov, B. V. A universal instability of many-dimensional
           oscillator systems. {\em  Phys. Rep.} {\bf 52}, 263 (1979).
\bibitem{lichtenberg} Lichtenberg, A. and Lieberman, M. Regular and chaotic 
dynamics. (Springer, N.Y., 1992).
\bibitem{divi} DiVincenzo, D. P. Quantum computation. {\em Science} {\bf 270},
255 (1995).
\bibitem{ekert} Ekert, A. and Josza, R. Quantum computation and Shor's 
factoring algorithm. {\em Review of Modern Physics} {\bf 68}, 733 (1996).
\bibitem{steane}  Steane, A. Quantum computing. 
                 {\it Rep. Progr. Phys.} {\bf 61}, 117 (1998).
\bibitem{kane} Kane, B.~E. A silicon-based nuclear spin quantum computer.
{\it Nature} {\bf 393}, 133 (1998).
\bibitem{ascat} Georgeot, B.  and Shepelyansky, D. L. Stable quantum 
    computation of unstable classical chaos. {\em Phys. Rev. Lett.} 
    {\bf 86}, 5393 (2001).
\bibitem{barrenco}Vedral, V., Barenco, A. and Ekert, A. Quantum networks
for elementary arithmetic operations.  
{\em Phys. Rev. A} {\bf 54}, 147 (1996).
\bibitem{pgm1}The quantum program can be written in the following form for
one map iteration, with C denoting a CNOT gate, T a Toffoli gate and R
a one-qubit rotation:

T(TCT)$^{n_q-1}$T$^{n_{q'}-n_q-1}$(CT)$^{n_{q'}-n_q-1}$CTCT
(CCTCT)$^{n_q-2}$CCTCRCC(TC)$^{n_{q'}-n_q-1}$
T$^{n_{q'}-n_q-1}$CT(TCT)$^{n_q-2}$
(CCTCT)$^{n_q-2}$CCTC

And for the time inversion:
R$^{n_{q'}}$CT$^{n_{q'}-2}$(CT)$^{n_{q'}-2}$\\
CCRT(TCT)$^{n_q-2}$(CCTCT)$^{n_q-2}$CCTC
\bibitem{maxwell} Maxwell, J. C. Theory of heat. (Longmass, Green and Co, 
                 London, 1871).
\bibitem{leff} Leff, H. S. and Rex, A.F. Maxwell's demon: entropy, 
              information, computing. (Adam Hilger, Bristol, 1990).
\bibitem{dima} Shepelyansky, D.~L.~ Some statistical properties of simple
classically stochastic quantum  systems. {\em Physica D} {\bf 8}, 208 (1983).
\end{thebibliography}
\end{document}